\documentclass[pra,twocolumn,showpacs,preprintnumbers,amsmath,amssymb,superscriptaddress]{revtex4}

\usepackage{color}
\usepackage{graphicx}
\usepackage{dcolumn}
\usepackage{bm}
\usepackage{amssymb}

\begin{document}
\title{Entanglement dynamics of two interacting qubits under the influence of local dissipation}
\author{ Muzaffar Qadir Lone}
\affiliation{
Department of Physics, University of Kashmir, Srinagar-190006,India}
\affiliation{
TCMP Division,
1/AF Salt Lake, Saha Institute of Nuclear physics, Kolkata, India.}


\date{\today}


\begin{abstract} 
We investigate the dynamics of entanglement given by the concurrence of a two-qubit system in 
the non-Markovian setting. A quantum master equation is derived which is solved in the eigen basis 
of the system Hamiltonian for $X$-type initial states. A closed formula for time 
evolution of concurrence is presented for a pure state. It is  shown  that under the influence of 
dissipation that non-zero entanglement is created in unentangled two qubit  states which decay in 
the same way as pure entangled states.  We also show that under real circumstances, the decay rate of 
concurrence is strongly modified by 
the  non-Markovianity  of the evolution.
\end{abstract}
\maketitle

\section{Introduction} 
The important ingredient for quantum computation and information processing is the 
presence of coherent superpositions. A single isolated two-level system can be 
prepared in a coherent superposition of  $|0\rangle$ and $|1\rangle$ states, and the 
manipulation of such states leads to new possibilities
for storage and processing of information \cite{Nielsen}.
In contrast to  the ideal isolated case, the interactions of real quantum systems with their 
environment lead to the loss of these coherent superpositions, in other words, decoherence. However, 
the more realistic case would be manipulation of many qubits. Coherent superposition of such states 
leads to the concept of entanglement, which forms a precious resource for quantum computation and 
information. The  fragility
of entanglement is due to the coupling between a quantum system and its environment;
such a coupling leads to decoherence, the process
 by which information is degraded \cite{MaxSc,zurek2}. 
In fact, decoherence is one of the main obstacles for the preparation, observation, and
implementation of multi-qubit entangled states.
The intensive work on quantum information and computing in recent years has 
tremendously increased the
interest in exploring and controlling decoherence effects   
\cite{nat1,milb2,QA,CJ,zurek,diehl,verst,weimer}.
In this work we address the problem where each of the two qubits are dissipatively coupled  to a 
local bosonic  bath; in quantum optical sense it would mean that  both the two-level 
systems are subject to spontaneous emission and  would imply there exist relaxation between the excited state to ground state.
Dissipation can assist the generation of entanglement 
\cite{MPl,Beige,PHoro} that can be used for various quantum information processing. For example, 
F. Verstraete {\it et al} \cite{verst} have shown that dissipation can be used as a resource for 
the universal quantum computation without any coherent dynamics needed to implement it. Contrary to
other  methods,  entanglement  generation  by  dissipation
does not require the preparation of a system in a particular input state and exists, in principle, for an arbitrary
long time.  These features
make dissipative methods inherently stable against weak
random perturbations, with the dissipative dynamics stabilizing the entanglement. 

The effects on system due to environment can be classified into the process with memory (Non-Markovian)
and without memory (Markovian) effects \citep{Pet,RF1,RF2,RF3,RF4,self}. In case of
Markovian processes, the environment acts as a sink for
the system information; the system of interest loses information into the environment and this lost information
plays no role in the dynamics of the system. However,
due to memory effects in case of non-Markovian dynamics, the information lost by the system during the interaction with the environment 
will return back to the system
at later time. This makes the non-Markovian dynamics
complicated. Understanding the nature
of non-Markovian dynamics is naturally a very important topic for quantum information science, where the aim is to control a quantum system
for use in technological applications \cite{Terhal,Ban,Wolf,Cirac}. In general, three time scales
in an open system exist to characterize non-Markovian
dynamics: (i) the time scale of the system (ii) the time
scale of the bath given by the bandwidth of bath spectral density (iii) the mutual time scale arising from the coupling between the system and the bath. It is usually
believed that non-Markovian effects strongly rely on the
relations among these different time scales \cite{27, 30, 31}.

In this paper we derive a quantum master equation for interacting qubits with local dissipation. The equation is derived utilizing the completeness of the 
eigen basis of the Hamiltonian representing the interacting qubits.  The time evolution of the density matrix turns out to be the sum of the time evolution corresponding 
to individual qubits with no cross terms. Next we solve this master equation for  $X$-type states with the assumption that individual baths have the same properties. The 
main content of this paper will remain however the same as for other kind of states and assuming different bath correlation functions for each bath. The different bath 
correlation functions can give rise to different time scales in the dynamics and is treated separately. Next we identify different regimes of dynamics 
(Markovian and non-Markovian) and  show that under non-Markovian regimes of the dynamics, there exists finite entanglement in an initially unentangled state.
This entanglement decay in the same way as the pure state entanglement and we find that the decay rate of entanglement is strongly modified by the non-Markovian behavior.

The rest of the paper is organized as follows:
In section II, we introduce the model Hamiltonian and derive the quantum master equation. In the past related works \cite{Pet,RF1,RF2,RF3,RF4}, non-interacting 
qubits have been considered. These qubits are then coupled to a common bath. However, in this paper, we consider qubits interacting through isotropic 
Heisenberg interaction which is a general kind of interaction in condensed matter physics. 
In section III, we solve the quantum master equation in the eigen-basis of the system Hamiltonian 
for a general class of initial quantum states under the assumption that the bath correlation 
function decay in the same way.In section IV we give the decay of entanglement of certain $X$-type 
state. Finally we conclude in section V with the remarks of wider context of our results.

\section{ Master equation for local dissipation}

In this section we will first derive master equation for the reduced density matrix of the 
system which  govern the dynamics of the system.
We consider two qubits represented by spin-$\frac{1}{2}$ particles or two level atoms coupled to 
each other via isotropic Heisenberg interaction. The qubits are subject to local dissipation 
through a coupling with a bosonic bath.
The Hamiltonian of the two qubit system is
\begin{eqnarray}
  H_s &=& J \overrightarrow{\sigma}_1 . \overrightarrow{\sigma}_2 \nonumber \\
 &=& J[\sigma_1^{+}\sigma_2^{-} + \sigma_1^{-}\sigma_2^{+} + \sigma_1^{z}\sigma_2^{z}],
\end{eqnarray}
where $\sigma_i^{\pm}= \frac{\sigma_i^x \pm i\sigma_i^y}{2}$. $J$ represents the energy scale of 
the system. The Hamiltonian $H_s$ can be diagonalized  exactly i.e.  $H_s|\psi_i\rangle = 
\epsilon_i |\psi_i\rangle$ where 
$|\psi_i\rangle$'s  are the eigenstates of the Hamiltonian $H_s$ with the eigen energies 
$\epsilon_i$ and are given below (with notation $|0\rangle=|\uparrow\rangle$ and 
$|1\rangle =|\downarrow\rangle$ ):

\begin{eqnarray}
 \epsilon_1= J;&& ~~~\psi_1 = |00\rangle \nonumber \\ 
\epsilon_2= 0; &&~~~\psi_2 = \frac{1}{\sqrt{2}}[|01\rangle + |01\rangle] \nonumber \\
\epsilon_3= -2J; &&~~~\psi_3 = \frac{1}{\sqrt{2}}[|01\rangle - |01\rangle] \nonumber \\
 \epsilon_4= J; &&~~~\psi_4 = |11\rangle. \nonumber 
 \end{eqnarray}

 We write the total Hamiltonian (system+bath) 
\begin{eqnarray}
 H=H_s + H_B + H_I
\end{eqnarray}
where $H_B$ is the Hamiltonian for the bath
\begin{eqnarray}
  H_B = \sum_{i=1}^{2} \sum_k \omega_k b_{i k}^{\dagger} b_{i k}, 
\end{eqnarray}
and the dissipative interaction of the system with bath is represented by the  Hamiltonian
\begin{eqnarray}
 H_I &=& \sum_{i=1}^{2} \sigma_i^x \sum_k[g_{i k} b_{i k} +g_{i k}^{\star} b_{i k}^{\dagger} ] 
\nonumber \\
&=& \sum_{i=1}^{2} \sigma_i^x  ( B_i + B_i^{\dagger})
\end{eqnarray}
where $B_i = \sum_k g_{i k} b_{i k}$.
Let $\tilde{O}(t)= e^{i H_o t} O e^{-i H_o t}$    represent an operator defined 
in interaction picture with respect to system and bath ($H_o= H_s +H_B$), we can therefore
write  $H_I$ in the interaction picture under rotating wave approximation as
\begin{eqnarray}
 \tilde{H}_I(t)=\sum_{i=1}^{2} [\tilde{\sigma}_i^{+}(t) \tilde{B}_i(t) + 
\tilde{\sigma}_i^{-}(t) \tilde{B}_i^{\dagger}(t)].
\end{eqnarray}
The time evolution of the system operators can be evaluated using the eigen basis of the system 
Hamiltonian $H_s$ as

\begin{eqnarray}
 \sigma_1^{+}(\tau) &=& \sum_{i,j=1}^{4} \!\!\! P_{ij} \langle \psi_i| \sigma_1^{+} \otimes 
I_2| \psi_j \rangle \exp[i (\epsilon_i-\epsilon_j)\tau] \\
\sigma_2^{+}(\tau) &=& \sum_{i,j=1}^{4} \!\!\! P_{ij} \langle \psi_i| I_1 \otimes \sigma_2^{+} 
| \psi_j \rangle \exp[i (\epsilon_i-\epsilon_j)\tau]
\end{eqnarray}
where $P_{ij} = |\psi_i\rangle \langle \psi_j |$ is the projection operator $P_{ij} P_{jk} = 
P_{ik}$ and $\sum_iP_{ii}=I$. In interaction picture, the time evolution of the total density 
matrix (system and bath) $\rho_T(t)$ is given by the 
von Neuman-Liouville equation as 

\begin{eqnarray}
 \frac{d \tilde{\rho}_T(t)}{dt} = -i[\tilde{H}_I(t), \tilde{\rho}_T(t)].
 \label{LOE}
\end{eqnarray}
Here we have used $\hbar=1$. We can formally integrate 
equation (\ref{LOE}) and write 
its solution as:

\begin{eqnarray}
 \tilde{\rho}_T(t)= \tilde{\rho}_T(0)- i \int_0^{t}\!\! ds  
 [\tilde{H}_I(s), \tilde{\rho}_T(s)].
\end{eqnarray}
Subsituting this solution back into the commutator of  equation (\ref{LOE}), we get upto second 
order 
the following equation:

\begin{eqnarray}
 \frac{d\tilde{\rho}_{T}(t)}{dt} =&-&i\left[ \tilde{H}_{I}\left(
t\right) ,\tilde{\rho}_{T}\left( 0\right) \right]  \nonumber \\
&-&\int\nolimits_{0}^{t}ds\left[ \tilde{H}%
_{I}\left( t\right) ,\left[ \tilde{H}_{I}\left( s\right) ,\tilde{%
\rho}_{T}\left( s \right) \right] \right].
\end{eqnarray}
The solution of the above equation depends on the initial conditions of total density operator. We 
consider an initially uncorrelated situation, i.e $\rho_T(0)= \rho_s(0) \otimes \rho_B$, where 
$\rho_s$ and $\rho_B$ are respectively the density operators for the system and bath. Tracing over 
the bath degrees of freedom and assuming that $tr_B[\tilde{H}_I(t) \rho_B]=0$, we get the following 
time non-local master equation for the reduced density matrix:

\begin{eqnarray}
 \frac{d \tilde{\rho}_s(t)}{dt}=- \int_0^{t} \!\!\! ds ~tr_B [ \tilde{H}_I(t), [ 
\tilde{H}_I (s), \tilde{\rho}_T(s)  ]    ].
\end{eqnarray}

As the bath degree of freedom are infinite so that the influence of the system on the bath is small
in the weak system-bath coupling case. As a cosequence, we write the total density operator $ 
\tilde{\rho}_T(s)= \tilde{\rho}_s(s) \otimes \rho_B + \mathcal{O}(\tilde{H_I})$ within the second 
order perturbation of system-bath coupling \cite{Pet,HJ,HFPB,HPB,MS,EF}. The 
replacement 
of total density matrix $\tilde{\rho}_T(s)$ with an  uncorrelated state $\tilde{\rho}_s(s) \otimes 
\rho_B $  is called as Born approximation. Therefore, under Born approximation we write

\begin{eqnarray}
 \frac{d \tilde{\rho}_s(t)}{dt}=- \int_0^{t}\!\!\! ds ~tr_B [ \tilde{H}_I(t), [ 
\tilde{H}_I (s), \tilde{\rho}_s(s) \otimes \rho_B ]    ].
\end{eqnarray}

The above equation is in a form of delayed integro-differential equation and is therefore a time 
non-local master equation. Replacing $\tilde{\rho}_s(s)$ with $\tilde{\rho}_s(t)$ in this equation 
\cite{Pet, HFPB, HPB} we get time-local master equation:

\begin{eqnarray}
 \frac{d \tilde{\rho}_s(t)}{dt}=- \int_0^{t}\!\!\! ds ~tr_B [ \tilde{H}_I(t), [ 
\tilde{H}_I (s), \tilde{\rho}_s(t) \otimes \rho_B ]    ].
\end{eqnarray}

Assuming the bath in the vacuum state initially, i.e $\rho_B=|0 \rangle \langle 0|$; using the form 
of  the $\tilde{H}_I(t)$, we arrive at the following equation:

\begin{eqnarray}
 \frac{d \tilde{\rho}_s(t)}{dt} = \mathcal{L}_1[\tilde{\rho}_s(t)] + 
\mathcal{L}_2[\tilde{\rho}_s(t)].
\end{eqnarray}
This forms  a non-trivial result. The master equations contains sums of $\mathcal{L}_i$ for each qubit and no cross terms with different $\mathcal{L}_i$'s.
 This result is the same as that for the non-interacting qubits. Here we have  ($i=1,2$)

\begin{eqnarray}
\!\!\! \!\!\! \!\!\! 
\mathcal{L}_i(\tilde{\rho}_s(t)) &=& \int_0^{t} \!\!\!  ds 
\{\Phi_i(t-s)[\tilde{\sigma}_i^{-} (s) 
\tilde{\rho}_s(t),\tilde{\sigma}_i^{+} (t) ] \nonumber \\
&&~~~~~~+ \Phi_i^{\dagger}(t-s) [\tilde{\sigma}_i^{-} (t), 
\tilde{\rho}_s(t) \tilde{\sigma}_i^{+} (s)] \}
\end{eqnarray}
and the bath correlation function is defined as
\begin{eqnarray}
 \Phi_i(t-s)&=& \langle B_i(t-s) B_i^{\dagger}\rangle_0 \nonumber \\
 &=&\sum_k |g_{i k}|^2 \exp[-i\omega_k (t-s)].
\end{eqnarray}
Next we revert  back to the Schr$\ddot{o}$dinger picture with a change in variable
$\tau=t-s$, we write

\begin{eqnarray}
\label{QME}
 \frac{d \rho_s(t)}{dt} &=& -i [H_s, \rho_s(t)] \nonumber \\
 &&+ \sum_{i=1}^{2} \int_0^{t} d\tau \left[\Phi_i(\tau)[\sigma_i^{-} (-\tau) 
\rho_s(t),\sigma_i^{+}  ] \right. \nonumber \\
&&~~~~~~~~~~~~~~\left. 
+ \Phi_i^{\dagger}(\tau) [\sigma_i^{-} , 
\rho_s(t) \sigma_i^{+} (-\tau)] \right].
\end{eqnarray}
This represents the quantum master equation in the Schr$\ddot{o}$dinger picture.
The solution of the above master equation depends on the type of initial states. In the next 
section we  find its solution for general $X$-type initial states.

 \section{Solution of Master equation}
 
 In order to obtain the dynamics of entanglement of our two qubit system, we assume that the qubits 
are initially  prepared in an X state \cite{TingYU}:

\begin{equation}
\rho_s(0)= \left [  
 \begin{array}{cccc}
 u(0) & 0 & 0 & w(0) \\
 0 & x_1(0) & y(0) & 0 \\
 0 & y^{\star}(0) & x_2(0) & 0 \\
 w^{\star}(0) & 0 & 0 & v(0)
 \end{array}
 \right]
\end{equation}
where we have used the standard basis $\{|00\rangle, |01\rangle, |10\rangle, |11\rangle\}$. Since 
the normalization and positivity of $\rho_s(0)$ i.e, 
$tr(\rho_s(0))=1$ and $\rho_s(0)>0$, the matrix elements $u, x_1, x_2, v$ are non-negative 
parameters with $u + x_1 + x_2 + v=1 $, $\sqrt{uv}\ge|w|$, and $\sqrt{x_1 x_2}\ge |y|$.
We can use more general  forms of density matrix with all elements non-zero, this makes the master equation intractable analytically. 
Next we express the X state $\rho_s(0)$  in the eigen basis of $H_s$ as

\begin{eqnarray}
 \rho_s(0) & =& a(0) |\psi_1 \rangle \langle \psi_1|  + b(0) |\psi_2\rangle \langle \psi_2| + e(0) 
|\psi_3\rangle \langle \psi_3| \nonumber \\
	    && + ~d(0) |\psi_4\rangle \langle \psi_4| + c(0) |\psi_1\rangle \langle \psi_4| + 
c^{\star}(0) |\psi_4\rangle \langle \psi_1| \nonumber \\
	  &&+~ h(0) |\psi_2\rangle \langle \psi_3| + h^{\star}(0) |\psi_3\rangle \langle \psi_2|
\end{eqnarray}
where the various parameters of the density operator in the eigen basis of $H_s$ are related to 
the parameters in the standard basis in the following way:
$a(0) = u(0)$,~$ b(0) =\frac{1}{2}[x_1(0) + x_2(0) + y(0) + y^{\star}(0)] $,~
$ e(0) = \frac{1}{2}[x_1(0) + x_2(0) - y(0) - y^{\star}(0)] $,~
 $h(0) = \frac{1}{2}[x_1(0) - x_2(0) - y(0) + y^{\star}(0)] $,
 $d(0)= v(0)$,~
 $c(0) = w(0)$.
 Next we see that the form of the density matrix is invariant during the time 
evolution generated by 
the quantum master equation. Therefore we can the density matrix at time $t$ as
\begin{eqnarray}
\label{rhoe}
 \rho_s(t) & =& a(t) |\psi_1 \rangle \langle \psi_1|  + b(t) |\psi_2\rangle \langle \psi_2| + e(t) 
|\psi_3\rangle \langle \psi_3| \nonumber \\
	    && + ~d(t) |\psi_4\rangle \langle \psi_4| + c(t) |\psi_1\rangle \langle \psi_4| + 
c^{\star}(t) |\psi_4\rangle \langle \psi_1| \nonumber \\
	  &&+~ h(t) |\psi_2\rangle \langle \psi_3| + h^{\star}(t) |\psi_3\rangle \langle \psi_2|.
\end{eqnarray}
In order to find the time evolution equations of the various parameters involved in equation 
(\ref{rhoe}) we   assume  that the bath correlation functions have the same  form 
\begin{eqnarray}
 \Phi_i(s) =\frac{\Gamma_i \lambda}{2} e^{-\lambda |s|}
\end{eqnarray}
where $\lambda$ is the spectral width of the bath, $\Gamma_i$ 
is related to the microscopic system-bath coupling constant. It defines the relaxation time scale 
$\tau_R$ over which the state of the system: $\tau_R \sim \Gamma_i ^{-1}$.
It can be shown to be related to the Markovian decay rate $\Gamma_M$  in Markovian Limit of flat 
spectrum. This form of correlation function corresponds to the Lorenztian spectral density of the 
bath \cite{Pet}. Assuming that $\Gamma_1= \Gamma_2 \equiv \Gamma_M$ for simplicity,  we substitute 
the $\rho_s(t)$ as in equation (\ref{rhoe}) in the quantum master equation  (\ref{QME})
and obtain the time dependence of the parameters as

\begin{eqnarray}
\label{f1}
  a(t) &=& a(0) e^{-\Gamma(t)} \\
  d(t) &= & d(0) + \int_0^{t}\!\!\!dz [\eta(z) b(z) + \Sigma(z) e(z)] \\
  c(t) &=& c(0) e^{-iS_1 t-\Gamma(t)} \\
  h(t) &=& h(0) e^{-iS_2 t - \Gamma(t)}
\end{eqnarray}

\begin{eqnarray}
\label{f2}
  \frac{d b(t)}{dt} + \eta(t) b(t)& =& \eta(t) a(t) \\
  \frac{d e(t)}{dt} + \Sigma(t) e(t) &= &\Sigma(t) a(t) 
\end{eqnarray}
where $\Gamma(t) =\Gamma_{+}(t) + \Gamma_{-}(t)$;
$\Gamma_{+}(t)=\frac{1}{2} \int_0^t \!dz~\Sigma(z) $;
$\Gamma_{-}(t)=\frac{1}{2} \int_0^t \!dz~\eta(z) $;
$S_1(t)= S_{+}(t) + S_{-}(t)$; $S_2(t)= 2Jt+S_1(t)$ and the explicit forms of these 
functions are given in the appendix A.

\section{Decay of entanglement}
In this section we study entanglement of a two qubit system by means of  concurrence 
\cite{Wooters}. For a density matrix $\rho$, concurrence  is defined as $\mathcal{C}=
max \{0,\sqrt{r_1}-\sqrt{r_2}-\sqrt{r_3}-\sqrt{r_4}\}$, where $r_1$, $r_2$, $r_3$ and  $r_4$ are 
the 
eigen values of matrix $R$ in the descending order. The matrix $R$ is defined as $R=\rho(\sigma^y_1 
\otimes  \sigma^y_2) \rho^{\star} (\sigma^y_1 \otimes  \sigma^y_2)  $ and  $\rho^{\star}$ represents
complex conjugation of $\rho$ in standard basis.
For X state in the standard 
basis we write concurrence as \cite{TingYU}

\begin{eqnarray}
 \mathcal{C}(t) \!= \!\! 2~ max\{0, |w(t)|-\!\sqrt{x_1(t) x_2(t)}, |y(t)|-\!\sqrt{u(t) v(t)}\} 
\nonumber\\
\end{eqnarray}
where we have
$  u(t)= a(t)$,~ $ w(t) = c(t)$,~
$ x_1(t) = \frac{1}{2}[b(t) + h(t) + h^{\star}(t) + e(t)]$,~
 $x_2(t)= \frac{1}{2}[b(t) - h(t) - h^{\star}(t) + e(t)]$,~
 $y(t) = \frac{1}{2}[b(t) - h(t) + h^{\star}(t) - e(t)]$,~
$ y^{\star}(t)= \frac{1}{2}[b(t) + h(t) - h^{\star}(t) - e(t)]$,~
$ v(t)= d(t)$. 
Next we use these results to investigate the decay of entanglement in some specific cases. First We 
consider  the decay of the  pure entangled state $|\Psi\rangle= \cos \frac{\theta}{2} |01\rangle + 
\sin\frac{\theta}{2}|10\rangle$. This  state has initial entanglement $\mathcal{C}(0)=\sin\theta$
and at time $t$ we write with the help of above results

\begin{eqnarray}
 \mathcal{C}(t) =  2~ max\{0, |y(t)|\}
\end{eqnarray}
or 
\begin{eqnarray}
 C(t) = 2|y(t)|=  \frac{1+\mathcal{C}(0)}{2} e^{-\Gamma_{-}(t)} -  \frac{1-\mathcal{C}(0)}{2} 
e^{-\Gamma_{+}(t)}.
\label{concu}
\end{eqnarray}
This forms an important result. It shows that even though we have initially an unentangled state 
$\mathcal{C}(0)=0$, we still have entanglement at later time $t$. This can be attributed to the 
dissipative interaction between the system and the bath. Let us suppose  $\theta =\pi$, which 
corresponds to $|10\rangle$ state; the effect of dissipative interaction ( $H_I(t)|10\rangle= 
B_1^{\dagger}(t)|11\rangle + B_2(t)|00\rangle$ ) results in  an entangled state.

\begin{figure}[]
\centering
 \includegraphics[width=1.5in,height=1.25in]{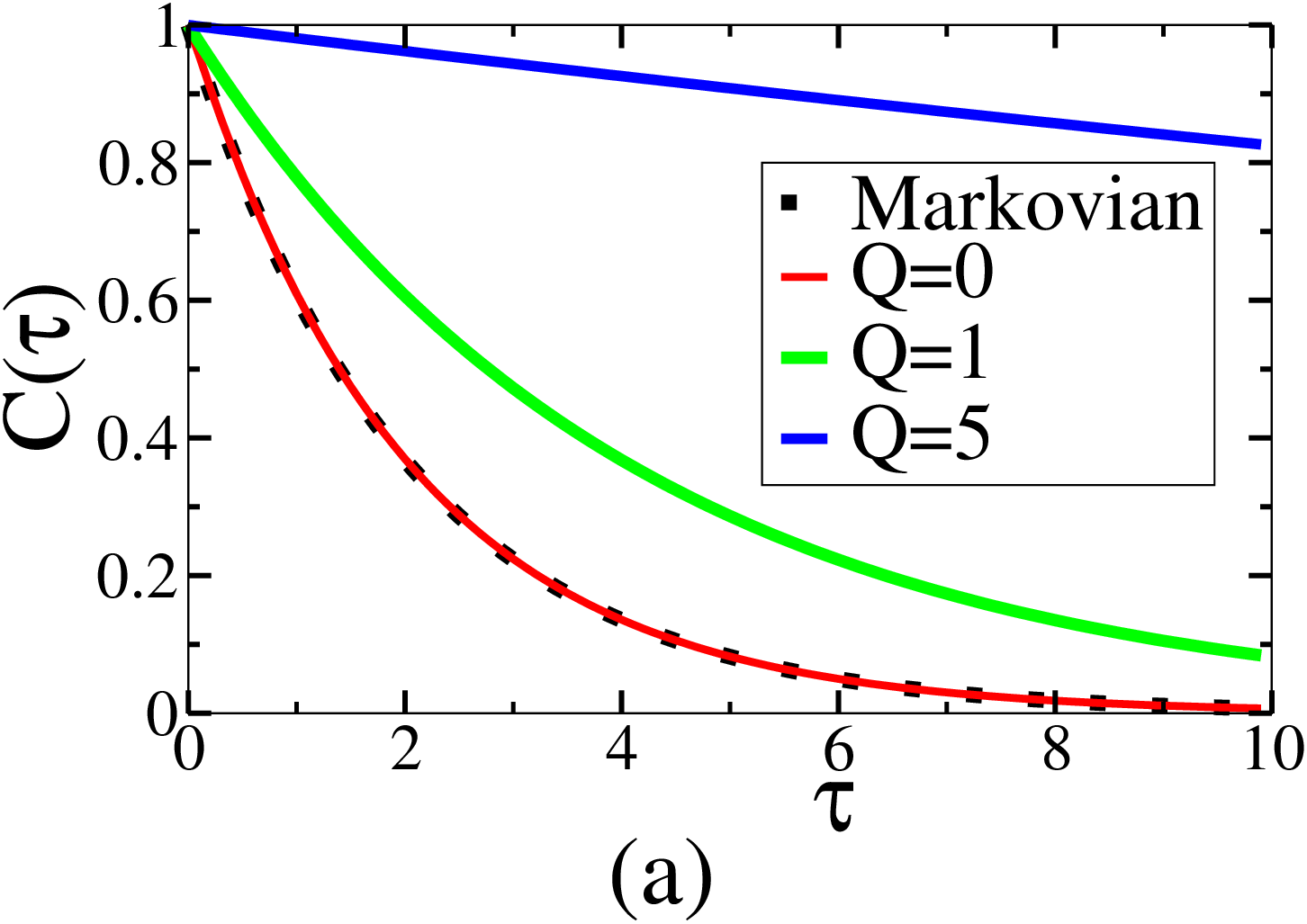}
 \hspace{0.35cm}
 \includegraphics[width=1.5in,height=1.25in]{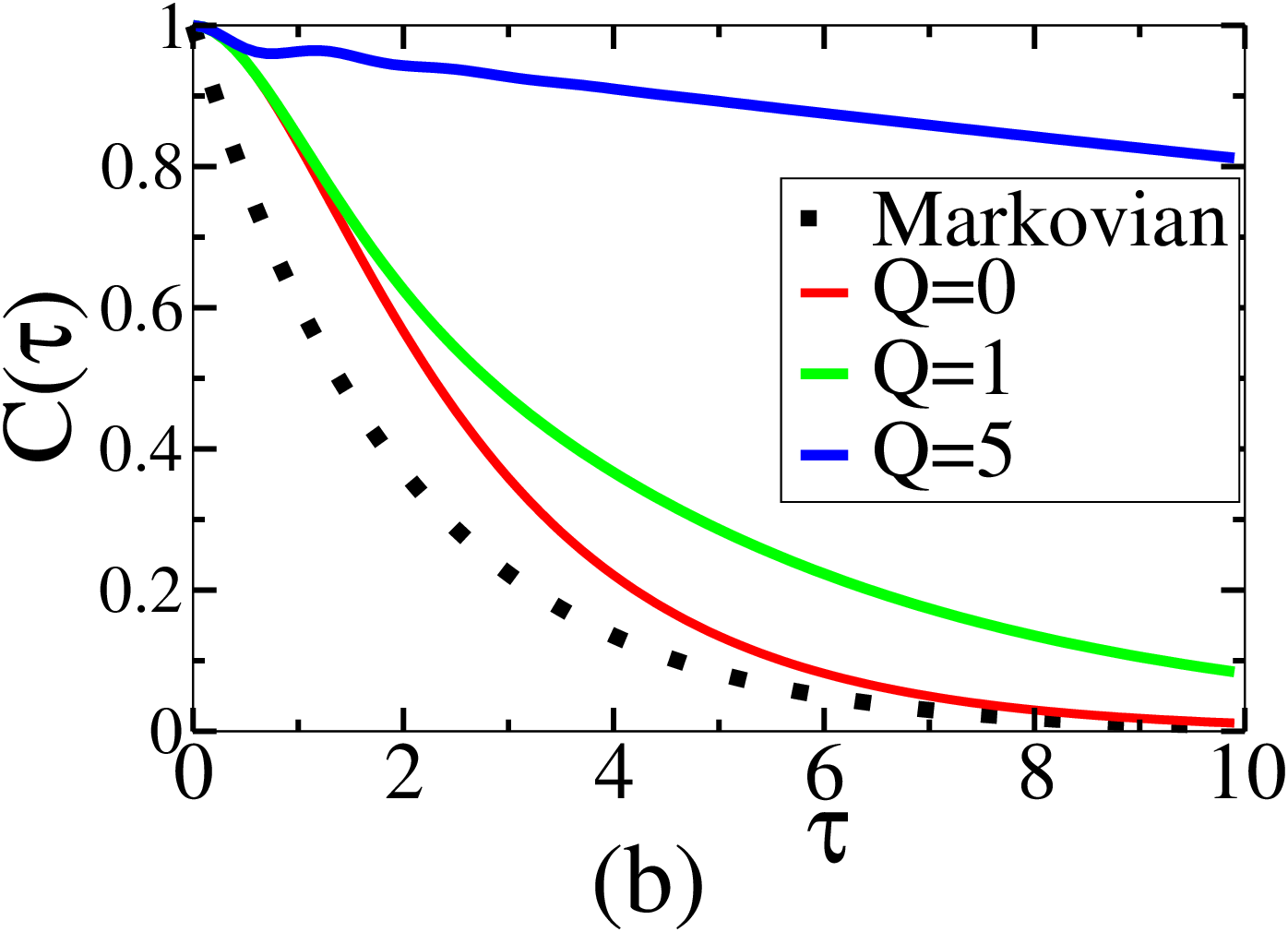}\\
 \vspace{0.61cm}
 \includegraphics[width=1.5in,height=1.25in]{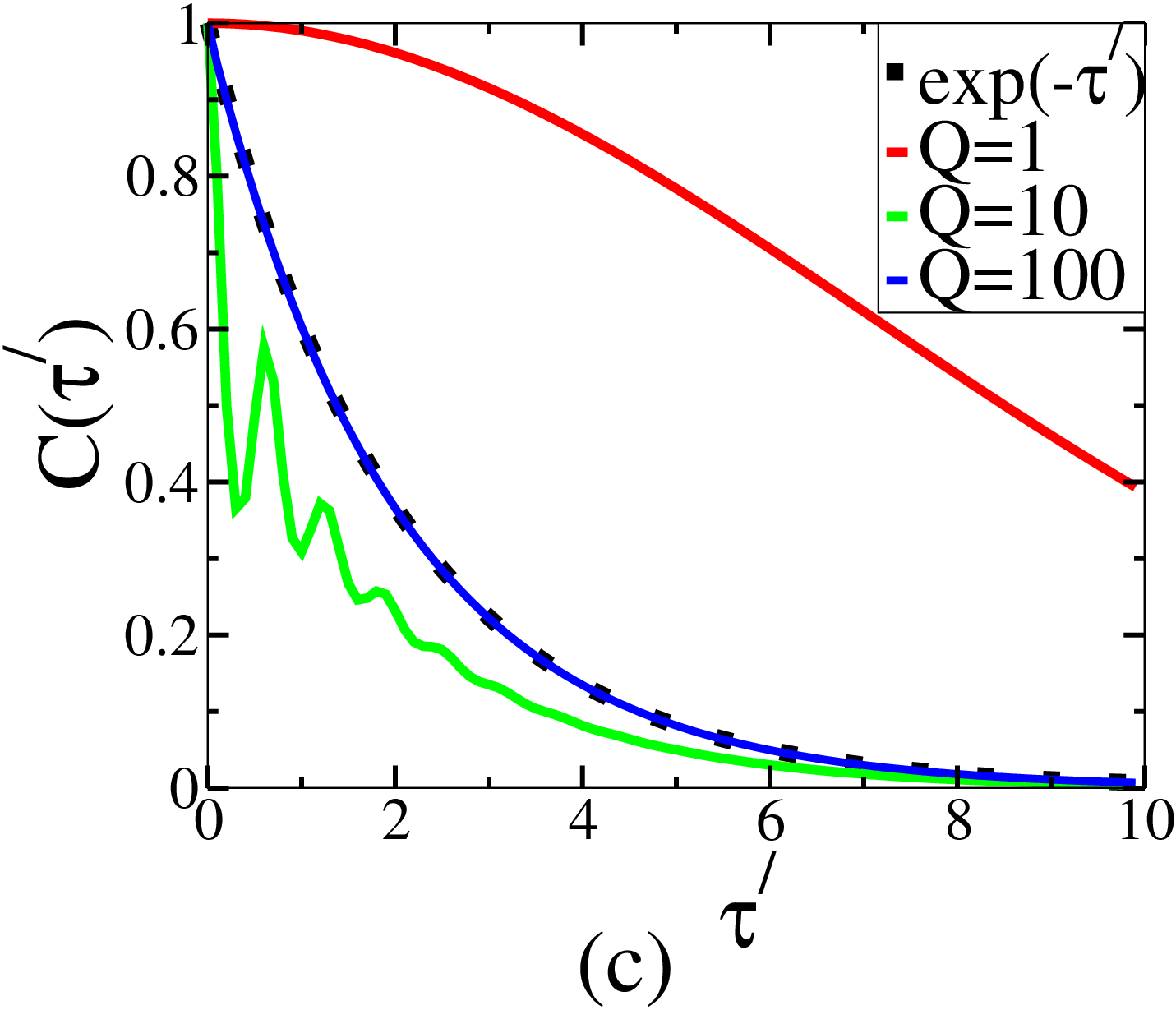}
 \hspace{0.35cm}
 \includegraphics[width=1.5in,height=1.25in]{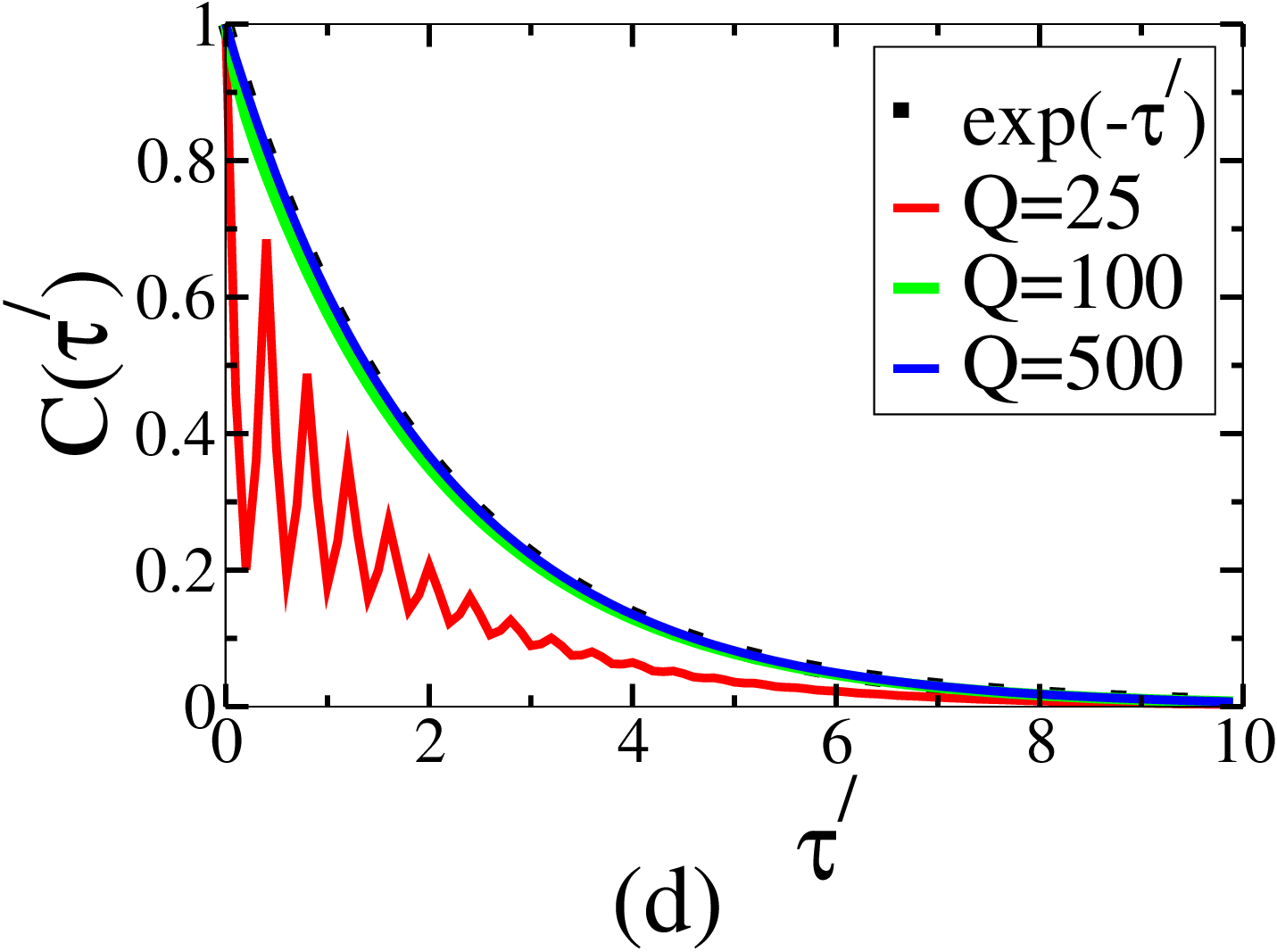}
\caption{Decay of entanglement as measured by concurrence $C(t)$ with time at 
different 
values of the parameter $Q$ and $R$. Here we have used $\theta=\pi/2$. Plots for different $Q$ 
values at (a) $R=100$ (b) $R=1$ (c) $R=0.01$ and (d) $R=0.001$. Plots (a) and (b) are in units of 
Markovian decay rate $\Gamma_M$, i.e $\tau=\Gamma_M t$, while plots (c) and (d) are in units of 
rescaled decay rate 
$\frac{\Gamma_M}{Q^2}$ i.e $\tau^{\prime}=\frac{\Gamma_M}{Q^2} t $ }.
\label{plot2}
\end{figure}

Next we analyze  the  Markovian and non-Markovian regimes of the dynamics, for that we  define the 
following parameters:
$  \tau=\Gamma_M t,~~~Q=\frac{J}{\lambda},~~R=\frac{\lambda}{\Gamma_M} $.
Therefore, using this parametrization we have


\begin{eqnarray}
 \Gamma_{+}(t) &=& \frac{\tau}{2[1 + 9Q^2]}    - \frac{[1-9Q^2][1-e^{-R\tau} 
\cos(3QR\tau)]}{2R[1+9Q^2]^2} \nonumber\\
 && -\frac{3Q}{R[1+9Q^2]^2}e^{-R\tau}\sin(3QR\tau) \\
 \Gamma_{-}(t)&=& \frac{\tau}{2[1 + Q^2]}   - \frac{[1-Q^2][1-e^{-R\tau} 
\cos(QR\tau)]}{2R[1+Q^2]^2} \nonumber\\
 && -\frac{Q}{R[1+Q^2]^2}e^{-R\tau}\sin(QR\tau).
\end{eqnarray}

In order to understand how the Markovian limit is obtained from the above expressions, we plot in 
Fig.~\ref{plot2}(a)-(b) $\mathcal{C}(\tau)$ for $\theta=\pi/2$ with respect to dimensionless 
parameter $\tau$ for 
$R=100$ and $R= 1 $ at different values of $Q$. We observe that the Markovian curve is recovered 
for $R>>1$ with $Q=0$. We can understand this behaviour of $\mathcal{C}(t)$ by looking at the 
different paramters involved. The typical time scale over which the system of two qubits changes is 
$\tau_s \sim 1/J$ and the time scale over which the bath changes is $\tau_B \sim 1/\lambda$ while 
relaxation time scale for each qubit would be given by $\tau_R \sim  1/\Gamma_M$. It means $R>>1$ 
implies $\tau_B<<\tau_R$ and $Q<1$ implies $\tau_R<\tau_s$. Thus physically $R>>1$ and $Q<1$ would 
imply that the system evolves over a large time compared to very fast bath dynamics. Therefore 
Markovian regime corresponds to $R>>1$ and $Q<1$ and we have  

\begin{eqnarray}
 \Gamma_{+}(t) = \Gamma_{-}(t)= \frac{\tau}{2}
\end{eqnarray}
and therefore we get standard Markovian limit: 
\begin{eqnarray}
 \mathcal{C}(t)=\mathcal{C}(0) e^{-\frac{1}{2}\Gamma_{M} t}.
\end{eqnarray}
We observe that under the Markovian limit an initially unentangled state remains unentangled 
always.  In situations where the spectral width $\lambda $ of the bath is narrower than the energy 
scale $J$ involved for the system  implying $Q>>1$. This would mean $\tau_R<<\tau_s$. In Fig. 
\ref{plot2}(b), we  observe  for $R=1$ that as $Q$ increases from 0, there is larger devaition from 
the Markovian dynamics of the concurrence $\mathcal{C}(t)$. The general trend is similar for all 
values of $ R $ as can be seen on comparison to Fig. \ref{plot2}(a). These observations suggest 
Markovian regime is in fact opposite to the regime $R<1$ and $Q>>1$, which we call as 
{\it non-Markovian} regime. The term that is responsible for this  larger deviation can be 
attributed to first terms of $\Gamma_{-}(t)$ and $\Gamma_{+}(t)$ containing $Q^2$ in the 
denominator. For $Q^2>>1$ suggest defining another time scale

\begin{eqnarray}
 \tau^{\prime} = \frac{\Gamma_M }{Q^2} t.
 \label{rescaled}
\end{eqnarray}

The decay of entanglement defined by $\mathcal{C}(t)$ at various values of $Q$ for non-Markovian 
regime $R<1$ in terms of rescaled time $\tau^{\prime} = \tau/Q^2$ is shown in the 
Fig.\ref{plot2}(c)-(d).   We see that for large $ Q $ the 
concurrence $\mathcal{C}(\tau^{\prime})$ coincides with exponential decay in units of the rescaled 
time. Next we see that before reaching the limiting behavior of exponential decay 
in rescaled time (\ref{rescaled}), we observe 
some oscillatory behavior Figure \ref{plot2}(c)-(d). The deviation from an exponential decay can be 
attributed to the memory effects developed in the two qubit system. This occurs clearly due to 
the second terms  in 
$\Gamma_{-}(t)$ and $\Gamma_{+}(t)$.  For $ Q \gg 1 $ we may approximate this as
\begin{eqnarray}
\Gamma_{-}(\tau') & \approx & \frac{1}{2}\left[
\tau'  + \frac{1- e^{-R Q^2 \tau' }\cos ( R Q^3  \tau')}{R 
Q^2}\right] \label{approxgamma1}\\
\Gamma_{+}(\tau') & \approx & \frac{1}{18}\left[
\tau'  + \frac{1- e^{-R Q^2 \tau' }\cos ( 3R Q^3  \tau')}{R 
Q^2}\right].
\label{approxgamma2}
\end{eqnarray}
%
 

In order for the oscillatory term to be visible, we require the exponential decay term in 
(\ref{approxgamma1})-(\ref{approxgamma2}) to be not too fast giving $ R Q^2 < 1 $, but 
simultaneously the oscillation 
frequency should be faster than the overall decay envelope $ R Q^3 > R Q^2 > 1 $.  The strongest 
oscillations therefore occur when $ R Q^2 \sim 1 $, which agree with the numerical plots in Figure 
\ref{plot2}(c)-(d). The deviation from an exponential decay can be attributed to the memory effects 
developed initially, typical of non-Markovian behavior.  The criterion for 
the strongest oscillatory behavior are satisfied when all the characteristic time scales  are all 
approximately the same i.e $\tau_R\sim \tau_s\sim \tau_B$.

\section{Conclusions}
In conclusion, we have derived a quantum master equation for system of two interacting qubits under 
the influence of local dissipation. Using the assumption that the  correlation functions have 
the same form for each of the baths, the solution of master equation is found
for the general $X$-type state. The time dependence of concurrence, a measure of 
entanglement is studied 
for a pure entangled state $|\Psi\rangle= \cos \frac{\theta}{2} |01\rangle + 
\sin\frac{\theta}{2}|10\rangle$ (a 
special case of $X$-type state) under both Markovian and non-Markovian regimes of the dynamics. It 
is found that for finite time  evolution an unentangled state can go to an entangled state 
in contrast to the Markovian case where the unentangled state remains unentangled always.
By identifying the parameter space, we have found that  our results reduce to 
standard Markovian decay rate which in general is not a physically relevant regime \cite{self}.
In the physically relevant regime with narrower spectral width as compared to $J$, the decay rate 
is better approximated by $\Gamma(t) =\frac{\Gamma_M}{Q^2} $, which is the standard Markovian decay 
rate divided by the $ Q^2 $, which can be quite large in practice.

Next we compare our work with the several other works that studied non-Markovian dynamics of entanglement. 
Taking the example of Ref. \cite{RF2}, the authors derive the non-Markovian decay of the entanglement  for the pure
state $|\Psi\rangle= \cos \frac{\theta}{2} |01\rangle + 
\sin\frac{\theta}{2}|10\rangle$. The time dependence of concurrence is given by: 

\begin{eqnarray}
 \mathcal{C}(t)= {\rm max} \{0, C(0) G(t)\}
\end{eqnarray}
where 
\begin{eqnarray}
  G(t)=e^{-\lambda t/2}\left[ \cosh(\frac{\lambda t}{2}\delta)+ \frac{1}{\delta}\sinh(\frac{\lambda t }{2}\delta) \right]
\end{eqnarray}
and  $\delta = \sqrt{1- \frac{2 \Gamma_M}{\lambda}}$ and $\Gamma_M$ is the Markovian decay 
rate. Our result Eqn. (\ref{concu}) is more general than above result. The results in these works \cite{RF1,RF2,RF3,RF4} do not tell about the 
amount of entanglement and its decay that would be present in an entangled state generated from dissipation.  
To see more clearly the behavior, let us examine this in two limiting cases

{\it Weak coupling limit} $ \Gamma_M \ll \lambda $: This regime corresponds 
to a weak coupling regime or a very broad coupling to many frequency modes, which gives Markovian behavior. 
Here $ \delta \approx 1 - \frac{\Gamma_M}{\lambda} $ and the decay function $G(t)$ gives purely exponential behavior.  
To first order the decay function may be approximated as
\begin{eqnarray}
 G(t) \approx e^{-\Gamma_M t/2}
\end{eqnarray}
which is nothing but standard Markovian spontaneous decay.  

{\it Strong coupling limit} $ \Gamma_M \gg \lambda $: The reverse regime is when the
linewidth of the bath  is extremely narrow, which gives rise to strongly non-Markovian behavior.  Here we may approximate 
$\delta =i\sqrt{\frac{2 \Gamma_M}{\lambda}} $ 
and 
\begin{eqnarray}
 G(t) = e^{-\lambda t /2}\left[
\cos \left( \sqrt{\frac{\Gamma_M \lambda}{2}} t \right)+ \sqrt{\frac{\lambda}{2 \Gamma_M}}\sin \left( \sqrt{\frac{\Gamma_M \lambda}{2}} t \right) 
\right] \nonumber \\
\end{eqnarray}
which corresponds to damped oscillations at frequency $\sqrt{\lambda \Gamma_M/2}$ and a decay 
envelope with rate $\lambda$. Thus we see that in both the cases the previous results do not yield the scaling factor $Q^2$ as 
derived in Eq. (35). 

The current result would be important for applications where spontaneous
emission is a serious drawback of using excited states, such as
for quantum information processors, quantum simulators, and
quantum metrological applications.

\appendix

\section{}
In this appendix we write the explicit forms of the various functions used in main text.

\begin{eqnarray}
\eta(t) &=& \frac{\Gamma_M \lambda}{\lambda^2 + J^2}\left[ 
\lambda (1-e^{\lambda t} \cos Jt) + Je^{-\lambda t}\sin Jt \right] \\
\Sigma(t)&=&\frac{\gamma_0 \lambda}{\lambda^2 +9 J^2}\left[ 
\lambda (1-e^{\lambda t} \cos 3Jt) + 3Je^{-\lambda t}\sin 3Jt \right] \nonumber \\
 \end{eqnarray}

\begin{eqnarray}
\Gamma_{-}(t) &=&\frac{1}{2} \int_0^t \!\!dz~\eta(z) \\
 &=&\frac{\Gamma_M \lambda}{2[\lambda^2 + J^2]}\left[ \lambda t 
 -\frac{\lambda^2-J^2}{\lambda^2+J^2}[1-e^{-\lambda t}\cos Jt)] \right.\nonumber \\
 && ~~~~~~~~~~~~~~~~~~\left. 
 - \frac{2\lambda J}{\lambda^2+J^2}e^{-\lambda t} \sin Jt \right]  \\
 \Gamma_{+}(t) &=& \frac{1}{2} \int_0^t \!\!dz~\Sigma(z) \\
  &=&\frac{\Gamma_M \lambda}{2[\lambda^2 + 9J^2]}\left[ \lambda t 
 -\frac{\lambda^2-9J^2}{\lambda^2+9J^2}[1-e^{-\lambda t}\cos 3Jt)] \right.\nonumber \\
 && ~~~~~~~~~~~~~~~~~~\left. 
 - \frac{6\lambda J}{\lambda^2+9J^2}e^{-\lambda t} \sin 3Jt \right] 
 \end{eqnarray}
 
 \begin{eqnarray}
 S_{-}(t) &=&\frac{\Gamma_M \lambda}{2[\lambda^2 + J^2]}\left[ J t 
 +\frac{2J}{\lambda^2+J^2}[1-e^{-\lambda t}\cos Jt)] \right.\nonumber \\
 && ~~~~~~~~~~~~~~~~~~\left. 
 + \frac{\lambda^2-J^2}{\lambda^2+J^2}e^{-\lambda t} \sin Jt \right]  \\ 
 S_{+}(t) &=&\frac{\Gamma_M \lambda}{2[\lambda^2 + 9J^2]}\left[ 3J t 
 +\frac{6J}{\lambda^2+9J^2}[1-e^{-\lambda t}\cos3 Jt)] \right.\nonumber \\
 && ~~~~~~~~~~~~~~~~~~\left. 
 + \frac{\lambda^2-9J^2}{\lambda^2+9J^2}e^{-\lambda t} \sin3 Jt \right]  
 \end{eqnarray}


\begin{thebibliography}{47}
\bibitem{Nielsen}   M. A.  Nielsen  and I. Chuang,   {\it Quantum Computation and Quantum
Communication} ( Cambridge University Press  ) $2000$.
\bibitem{MaxSc} M. Schlosshauer, Rev. Mod. Phys. {\bf 76}, 1267 $2005$
\bibitem{zurek2}W. H.  Zurek,    Rev. Mod. Phys. {\bf 75}, 715(2003) . 
\bibitem{nat1} J. T. Barreiro, P. Schindler , O. G\"{u}hne ,T.  Monz , M.  Chwalla ,
  C. F.  Roos  , M. Hennrich  and R. Blatt,    Nature Phys. {\bf 6}, 943 (2010). 
\bibitem{milb2} S.  Schneider   and  G. J. Milburn,  Phys. Rev. A {\bf 57}, 3748 (1998).
\bibitem{QA} Q. A. Turchette,  C. J.  Myatt, B. E.  King , C. A. Sackett, D. Kielpinski ,
W. M. Itano , C. Monroe  and D. J.  Wineland,  Phys. Rev. A, {\bf 62}, 053807 (2000).
\bibitem{CJ} C. J.  Myatt, B. E.  King , Q. A. Turchette, C. A. Sackett, D. Kielpinski ,
W. M. Itano , C. Monroe  and D. J.  Wineland     Nature, {\bf 403}, 269 (2000). 
\bibitem{zurek}  E. Knill , R. Laflamme  and W. H. Zurek, Science {\bf 279}, 342 (1998).
\bibitem{diehl} S.  Diehl , A. Micheli, A. Kantian, B. Kraus, H. Buechler, P. Zoller,   Nature 
Phys. {\bf 4}, 878 (2008). 
\bibitem{verst} F.Verstraete, M. M.Wolf, and J. I.Cirac, Nat. Phys.{\bf 5}, 633 (2009).
\bibitem{weimer} H. Weimer , M. M\"{u}ller  , I. Lesanovsky , P. Zoller  and H.P.  B\"{u}chler, Nature Phys. {\bf 6}, 382 (2010).
\bibitem{MPl} M. B. Plenio et al., Phys. Rev. A 59, 2468 (1999).
\bibitem{Beige} Beige et al., J. Mod. Opt. 47, 2583 (2000).
\bibitem{PHoro}P. Horodecki, Phys. Rev. A 63, 022108 (2001).
%
\bibitem{Pet}  H. P. Beuer  and F. Petruccione,    {\it The Theory of Open 
Quantum systems} (Oxford, New York: Oxford University Press)(2005).
\bibitem{RF1}  R. Lo Franco, B. Bellomo, S. Maniscalco, and G. Compagno, Int. J. Mod. Phys. B {\bf 27}, 1345053 (2013).
\bibitem{RF2} B. Bellomo, R. Lo Franco, and G. Compagno, Phys. Rev. Lett. {\bf 99}, 160502 (2007).
\bibitem{RF3} B. Bellomo, R. Lo Franco, and G. Compagno, Phys. Rev. A {\bf 77}, 032342 (2008).
\bibitem{RF4} K. M. Fonseca Romero and R. Lo Franco, Phys. Scr. {\bf 86}, 065004 (2012).
\bibitem{self} M. Q. Lone, and T. Byrnes, Phys. Rev. A {\bf 92}, 011401({\bf R}),  (2015).
\bibitem{Terhal} B. M. Terhal and G.Burkard, Phys. Rev. A 71, 012336
(2005).
\bibitem{Ban} M. Ban, J. Phys. A: Math. Gen. 39, 1927 (2006).
\bibitem{Wolf} M. M. Wolf, J. Eisert, T. S. Cubitt, and J. I. Cirac, Phys.
Rev. Lett. 101, 150402 (2008).
\bibitem{Cirac} F. Pastawski, L. Clemente and J. I. Cirac, Phys. Rev. A
83, 012304 (2011).
\bibitem{27} P. Haikka and S. Maniscalco, Phys. Rev. A 81, 052103
(2010).
\bibitem{30} W. M. Zhang, P. Y. Lo, H. N. Xiong, M. W. Y. Tu, and
F. Nori, Phys. Rev. Lett. 109, 170402 (2012).
\bibitem{31} M. A. Cirone, G. De. Chiara, G. M. Palma and A. Recati,
New J. Phys. 11 103055 (2009).
\bibitem{HJ} H. J.  Carmichael, {\it Statistical Methods in Quantum Optics I } 
(Berlin: Springer-Verlag) (2008).
\bibitem{HFPB}   H. P. Breuer, B. Kappler and F. Petruccione, Phys. Rev. A {\bf 59}, 1633 (1999). 
\bibitem{HPB} H. P. Breuer, B. Kappler and F. Petruccione ,  Ann. Phys. (NY)  {\bf 291}, 36 (2001). 
\bibitem{MS} M. Schr\"{o}der , U. Kleinekath\"{o}fer  and M.  Schreiber,   J. Chem. Phys. {\bf 
124} 084903 (2006).
\bibitem{EF}  E. Ferraro ,M. Scala, R. Migliore  and A. Napoli,  Phys. Rev. A {\bf 80}, 042112 (2009).
\bibitem{TingYU} T. Yu and J. H. Eberly, Quantum Inf. Comput. 7, 459 (2007)
\bibitem{Wooters} W.K.Wooters, Phys. Rev.Lett. {\bf 80}, 2245 (1998).
\end{thebibliography}
\end{document}